\documentclass[preprint,aps,prd,amssymb,showpacs,superscriptaddress,nofootinbib]{revtex4}
\usepackage{graphicx}

\newcommand{\Case}[2]{{\textstyle \frac{#1}{#2}}}
\newcommand{\lP}{\ell_{\mathrm P}}

\begin{document}

%\today
\preprint{IMSc/2008/11/15}

\title{Effective Actions from Loop Quantum Cosmology: Correspondence
with Higher Curvature Gravity}

\author{Ghanashyam Date}
\email{shyam@imsc.res.in}
\affiliation{The Institute of Mathematical Sciences\\
CIT Campus, Chennai-600 113, INDIA.}
\author{Sandipan Sengupta}
\email{sandi@imsc.res.in}
\affiliation{The Institute of Mathematical Sciences\\
CIT Campus, Chennai-600 113, INDIA.}

\begin{abstract}
Quantum corrections of certain types and relevant in certain regimes can
be summarised in terms of an effective action calculable, in principle,
from the underlying theory. The demands of symmetries, local form of
terms and dimensional considerations limit the form of the effective
action to a great extent leaving only the numerical coefficients to
distinguish different underlying theories.  The effective action can be
restricted to particular symmetry sectors to obtain the corresponding,
{\em reduced effective action}. Alternatively, one can also quantize a
classically (symmetry) reduced theory and obtain the corresponding
effective action. These two effective actions can be compared.  As an
example, we compare the effective action(s) known in isotropic loop
quantum cosmology with the Lovelock actions, as well as with more
general actions, specialized to homogeneous isotropic space-times and
find that the $\bar{\mu}$-scheme is singled out. 
\end{abstract}

\pacs{04.60.Pp, 04.60.Kz, 98.80.Jk}

\maketitle

%%%%%%%%%%%%%%%%%%%%%%%%%%%%%%%%%%%%%%%%%%%%%%%%%%%%%%%%%%%%%%%%%%%%%%%
\section{Introduction}
It is quite common to incorporate various types of quantum corrections
in an {\em effective action} which contains the classical action one
begins with. The degree of generality chosen for the form of an
effective action, reflects the context of its proposal/computation and
usually accounts for a class of quantum corrections. For example, the
effective action formally defined from the path integral is in general
expected to be non-local. However, in perturbation theory it is
typically obtained as an infinite power series each term of which is
{\em local} in the basic fields and their derivatives. It incorporates
the perturbative quantum corrections. Such effective actions can be
constructed starting from any classical action, in particular for both a
``full theory'' and its ``reduced versions'' corresponding to some
chosen (classical) sectors thereof. The same reduction procedure can be
carried out for the full theory effective action, possibly with further
approximations. Thus we have two effective actions and a comparison is
conceivable. Such comparisons could shed some light on
quantize-after-reduction and reduce-after-quantization approaches.
However, if one has only effective actions for different classical
sectors of a theory, then the demand that these be obtainable from a
common full theory effective action could be used to constrain some of
the quantization ambiguities of the reduced models.

We attempt such an exercise in the context of the effective actions
available in the loop quantization of the homogeneous and isotropic
sector of Einstein's theory. Assuming the usual higher derivative
effective actions for the full theory, it follows that of the various
quantization schemes available in isotropic loop quantum cosmology
(LQC), the so-called $\bar{\mu}$-scheme is the most natural one. The
scope and limitations of such comparisons is also discussed.

The paper is organized as follows. In section II, we take the effective
Hamiltonian from LQC and obtain a corresponding Lagrangian in a suitable
form.  These are written in a form which subsumes the old
$\mu_0$-scheme, the improved $\bar{\mu}$-scheme as well as the form
coming from lattice refinement models. In section III, we consider the
Lovelock action in arbitrary space-time dimensions, specialize them for
the FRW form of metric and compare with the LQC effective actions. In
the next section, we discuss the more general forms of actions, in four
space-time dimensions, specialise to the FRW metric and discuss methods
for comparing with the LQC effective action. The final section contains
a summary and some remarks.

\section{Effective Action from LQC}

Let us consider the simplest of the homogeneous models, namely the
isotropic model suitably coupled to scalar matter. The fundamental
discreteness of the LQC implies two distinct types of corrections: (a)
those arising from replacement of connections by holonomies and (b)
those arising due to the unusual definition of the inverse volume
operator forced by the discreteness. The latter ones typically show up
in the matter sector and are absent in the gravitational part for the
spatially flat, isotropic models. The modifications implied by these are
small deviations in the classical regime: $\lP\Case{\dot{a}}{a} \ll 1$.
In this regime, the corrections are summarised in an {\em effective
Hamiltonian} which has been obtained from the quantum Hamiltonian
constraint using a {\em leading order} WKB approximation
\cite{BanerjeeDate} or via expectation values in the (kinematical)
coherent states \cite{Willis}.  In the gravitational sector it takes the
form,
\begin{eqnarray} 
H_{\text{grav}} & = & - \frac{3}{\kappa}\sqrt{p}\left[\frac{
\text{sin}^2\left(\epsilon(\alpha, p) K\right)}{\epsilon^2(\alpha,
p)}\right] ~ ~,~ ~ \{K, p\} ~ = ~ \frac{\kappa}{3} ~ , ~ \kappa := 8 \pi
G \ , \nonumber \\
\epsilon(\alpha, p) & := & \mu(\alpha, p) \gamma ~ ~ , ~ ~ \mu(\alpha,
p) :=
\mu_{\alpha}\left(\frac{\sqrt{\gamma}\lP}{\sqrt{p}}\right)^{-2\alpha} \
.  
\end{eqnarray}
The $K$ is related to the usual `connection variable' $c$ by $K := c
\gamma^{-1}$. The parameter $\alpha$ is used to denote various
quantization schemes. $\mu_{\alpha}$ is a constant, $\gamma$ is the
Barbero-Immirzi parameter and $\lP$ is the Planck length (its precise
definition does not matter for our purpose).

The older quantization scheme \cite{ABL} is obtained for $\alpha = 0$
with $\mu(0, p) := \mu_0$, a fixed ambiguity parameter. The improved
quantization of \cite{APS} is obtained for $\alpha = - \Case{1}{2}$ with
$\mu_{-1/2} := \Delta$ while \cite{Martin} permits all values of
$\alpha$, $ -1/2 \leq \alpha \leq 0$. Note that the classical
Hamiltonian is recovered in the limit $\epsilon(\alpha, p) \to 0$.

We will first make a simple ``canonical'' transformation so that the
scale factor is the configuration space variable and then do an inverse
Legendre transformation to get the corresponding Lagrangian which is a
function of only the scale factor and its time derivative. This is the
form that we will compare with a general form of an effective action
specialized to the FRW metric.

We begin by introducing the identification $a := \xi \sqrt{p}$ and its
conjugate variable $p_a(K, p)$ to be chosen such that $\{a, p_a\} =
\Case{\kappa}{3}$. This leads to a choice, $p_a(K, p) := -
\Case{2\sqrt{p}}{\xi}K$. Substituting $\sqrt{p} = a\xi^{-1}, \ K = -
\Case{\xi^2}{2}\Case{p_a}{a}$ in the Hamiltonian leads to,
\begin{equation}
H_{\text{grav}}(a, p_a) ~ = ~ -
\frac{3}{\kappa}\frac{a}{\xi}\frac{1}{\epsilon^2(\alpha, a)}
\text{sin}^2\left(\frac{\epsilon(\alpha, a) \xi^2 p_a}{2 a}\right) ~ ~,~ ~ \{a,
p_a\} ~ = ~ \frac{\kappa}{3} \ .  
\end{equation}

For the synchronous time (lapse equal to 1), one gets,
\begin{equation} \label{VelocityEqn}
\dot{a} ~ = ~ \{a, H_{\text{grav}}(a, p_a)\} ~ = ~ -\frac{\xi}{2
\epsilon(\alpha, a)}\mathrm{sin}\left(\frac{\epsilon(\alpha, a)\xi^2p_a}{a}\right) \ .
\end{equation}

The Lagrangian is obtained by inverse Legendre transformation,
\begin{equation}
L(a, \dot{a}) ~ := ~ \frac{3}{\kappa} \dot{a} p_a(a, \dot{a}) -
H_{\text{grav}}(a, p_a(a, \dot{a}) )  \ ,
\end{equation}
where $p_a(a, \dot{a})$ is to be obtained by inverting
(\ref{VelocityEqn}). This is easily done and gives,
\begin{eqnarray}
\mathrm{sin}^2\left(\frac{\epsilon(\alpha, a)\xi^2 p_a}{2a}\right) & = & \frac{1
- \sqrt{1 - \Case{4\epsilon^2(\alpha, a)\dot{a}^2}{\xi^2}}}{2} \\
H_{\text{grav}}(a, \dot{a}) & = & -
\frac{3}{2\kappa}\frac{a}{\xi}\frac{1}{\epsilon^2(\alpha, a)} \left[1 - \sqrt{1
- x^2} \right] ~ ~ , ~ ~ x ~ := ~ \Case{2\epsilon(\alpha, a)\dot{a}}{\xi}  \\
L(a, \dot{a}) & = & \left[\frac{3}{\kappa}\right] \dot{a}\left\{ -
\frac{a}{\epsilon(\alpha, a)\xi^2} \mathrm{sin}^{-1}(x) \right\} -
H_{\text{grav}}(a, \dot{a}) \nonumber \\
& = & -
\left[\frac{3}{2\kappa}\right]\left[\frac{a}{\xi\epsilon^2(\alpha, a)}\right]\left[
x \ \mathrm{sin}^{-1}x - 1 + \sqrt{1 - x^2}\right] \ .
\end{eqnarray}

The third bracket can be expressed as a power series in $x$ as,
\begin{equation}
x \ \mathrm{sin}^{-1}x - 1 + \sqrt{1 - x^2} ~ = ~ \sum_{n = 1}^{\infty}
\frac{x^{2n}}{n! 2^n}\frac{(2n - 3)!!}{(2n - 1)} ~,~ 
\end{equation}
where $n!! := 1\cdot 3 \cdot 5 \cdots [n/2]$ and equals 1 for $n = 0$. 

Observe that the $\epsilon(\alpha, a)^{-2}$ factor cancels, leading to
the Lagrangian, 
\begin{eqnarray}
L(a, \dot{a}) & = & - \left[\frac{3}{\kappa}\frac{a
\dot{a}^2}{\xi^3}\right]\left[1 + \sum_{n = 1}^{\infty} \frac{x^{2n}}{(n
+ 1)! 2^n}\frac{(2n - 1)!!}{(2n + 1)} \right]  ~ ~,~ ~
\label{EffectiveLagrangian}\\ 
x & = & \left\{ \begin{array}{lcr} 2 \mu_0 \gamma \xi^{-1} \dot{a} &
\hspace{3.0cm} & (\mu_0-\mathrm{quantization})\\
2 \Delta \gamma^{3/2}\Case{\lP}{a}\dot{a} & \hspace{3.0cm} &
(\bar{\mu}-\mathrm{quantization}) \\
2 \mu_{\alpha} \gamma
\left(\frac{\xi\sqrt{\gamma}\lP}{a}\right)^{-2\alpha} \xi^{-1}\dot{a} &
\hspace{3.0cm} & (\mbox{lattice refinement}) 
\end{array} \right. \ . \nonumber
\end{eqnarray}

For comparison with the classical theory, we recall that for the FRW
spatially flat metric, one has
\begin{eqnarray}
ds^2 & := & dt^2 - a^2(t)\{dx^2 + dy^2 + dz^2\} \nonumber \\
R & = & -6 \frac{\ddot{a}}{a} - 6 \frac{\dot{a}^2}{a^2} ~ ,  \nonumber
\\
S & := & - \frac{1}{16\pi G}\int dt\int_{\mathrm{cell}} d^3x \sqrt{|det
g|} R  ~ ~ = ~ ~ - \frac{3 V_0}{\kappa} \int dt \ a \dot{a}^2 \ .
\label{ClassicalLagrangian}
\end{eqnarray}
Here $V_0$ is the comoving volume of a fiducial cell necessary in an
action formulation.  Comparison of the first term of
(\ref{EffectiveLagrangian}) and (\ref{ClassicalLagrangian}) suggests the
identification $\xi = V_0^{-1/3}$. 

Notice that only for $\alpha = - 1/2$, does the dependence on the
fiducial cell (through $\xi$) disappear and the Lagrangian becomes a
power series in $\lP\Case{\dot{a}}{a}$.  In all cases, the degrees of
freedom remain exactly the same and only the specification of the
dynamics deviates from the Einsteinian one.

Now the question which is raised many times is whether the loopy quantum
corrections that have been summarised in the LQC effective actions
above, are ``analogous'' to the higher derivative terms expected in an
effective action for the full theory. One way to explore this question
is to look for an effective action for full theory, restrict it to the
homogeneous and isotropic sector and compare with the LQC effective
action(s). If an action for the full theory continues to lead to a
second order (in time) field equation, then the degrees of freedom
remain the same as those implied by the Einstein-Hilbert action and on
restriction to the FRW sector, the same feature will continue to hold.
Such actions are indeed available and are known as the Lovelock actions.
We discuss these and their reduction, in the next section.

%%%%%%%%%%%%%%%%%%%%%%%%%%%%%%%%%%%%%%%%%%%%%%%%%%%%%%%%%%
\section{Lovelock Actions}
There is a special class of actions involving homogeneous polynomials in
the Riemann tensor, Ricci tensor and Ricci scalar which have the
property that the corresponding equations of motion are second order in
time. These are the Lovelock actions \cite{Lovelock}. If these are
specialised to the FRW metrics, then they become a function of only $a,
\dot{a}$ (up to total time derivatives).  In the following, the
space-time is taken to be $D$ dimensional. 

The pure Lovelock terms, $L_n$, are $2n$-th order homogeneous
polynomials in the $R_{abcd}, R_{ab}$ and $R$, with coefficients chosen
so as to remove higher derivative terms. Explicitly \footnote{Here we
follow the same notation as in \cite{Deruelle}.},
\begin{equation}
L^D_{n} = \frac{1}{2^n}~\delta ^{a_{1}...a_{2n}}
_{b_{1}...b_{2n}}~R^{~~~~b_{1} b_{2}}_{a_{1} a_{2}} \cdots
R^{~~~~b_{2n-1}b_{2n}}_{a_{2n-1}a_{2n}}, 
\end{equation} 
where $~\mathrm{\delta^{a_{1}...a_{2n}}_{b_{1}...b_{2n}}}~$ is the
Kronecker symbol of order $2n$ (totally antisymmetric in both sets of
indices) and $~\mathrm{R_{abcd}}~$ is the D-dimensional Riemann tensor.
One may notice that $L^D_0 = 1$ corresponds to the cosmological constant
term while the $L^D_1 = R $ is the familiar Einstein Hilbert term. Next
comes the Gauss-Bonnet term 
\begin{equation} 
L^D_2~=~R^2~-~4~R_{ab}R^{ab}~+~R_{abcd}R^{abcd} .  
\end{equation} 
and so on. The $L^D_n$ has dimensions of (length)$^{-2n}$. 

Due to the antisymmetrization, in D-dimensions, all Lovelock terms with
$\mathrm{n >\frac{D}{2}}$ vanish identically. For even D, the
$L^D_{D/2}$ is a total derivative, and thus doesn't contribute to the
equations of motion. For even D the $\sqrt{|g|} L^D_{D/2}$ is in fact
the Euler density, a topological invariant for the manifold.  The
Lovelock action in $D$ dimensions is a {\em linear combination} of the
non-vanishing pure Lovelock terms. To facilitate comparison with the
LQC, we drop the cosmological constant term. 

Consider now the FRW metric in $D$ space-time dimensions.
\begin{eqnarray}
ds^2 &~ =~ & dt^2~-~a^2(t)~\left[\frac{dr^2}{1 - kr^2} + r^2
d\chi_{1}^2~+~r^2 sin^{2}\chi_{1} d\chi_{2}^2~+~...\right]\nonumber\\
& ~=~ & dt^2~-~a^2(t)~\left[\frac{dr^2}{1-kr^2}~+~r^2
d\Omega_{D-2}^2\right] ~ ~\mathrm{where,}
\end{eqnarray}
$~a(t)~$ is the scale factor, $r$ is the radial coordinate while
$\chi_i$ are the angular coordinates on the $(D -2)$ dimensional sphere.
We consider the spatially flat case ($k=~0$) throughout. 

The non-zero components of the Riemann and the Ricci tensors are given
by, 
\begin{eqnarray}
\mathrm{Riemann:} & & \nonumber \\
& & R^{tr}_{~ ~tr} ~ = ~ R^{t \chi_i}_{~ ~ ~t \chi_i} ~ = ~
-\frac{\ddot{a}}{a} ~ ~ ~ ~ , ~ ~ 
R^{r \chi_i}_{~ ~ ~r \chi_{i}} ~=~ - \frac{\dot{a}^2}{a^2} ~ ~ , ~ ~ i =
1, \cdots, D - 2 \ .  \label{Riemann}\\
& & R^{\chi_{i} \chi_{i+j}}_{\hspace{1.0cm}\chi_{i}\chi_{i+j}} ~ = ~ -
\frac{\dot{a}^2}{a^2}~ ~ ~ , ~ ~ i = 1, \cdots, D - 2 ~ , ~ j = 1,
\cdots D - 2 - i \ .\nonumber \\
\mathrm{Ricci:} \hspace{0.7cm} & & \nonumber \\
& & R^r_{~r} ~ = ~ - \frac{\ddot{a}}{a} - (D-2)\frac{\dot{a}^2}{a^2} ~ ~
, ~ ~
R^t_{~t} ~ = ~ -~(D-1)\frac{\ddot{a}}{a} ~ ~ , ~ ~  \nonumber \\
& & R^{\chi_{i}}_{~ ~\chi_{i}} ~ = ~ R^r_{~r} ~ ~ , ~ ~ i = 1, \cdots, D
- 2 \ .  \label{RicciTensor}
\end{eqnarray}

The Ricci scalar is given by
\begin{equation} \label{RicciScalar}
R ~ = ~ -
(D-1)~\left[2~\frac{\ddot{a}}{a}~+~(D-2)\frac{\dot{a}^2}{a^2}~\right]
\end{equation}

Using these expressions we obtain\footnote{Explicit expressions of
$~L_2, L_3~$etc. in terms of the curvature invariants can be found in
\cite{MullerHoissen}.},
\begin{equation}
\sqrt{|g|}L^D_n ~ = ~ a^{D -1} \left[-\frac{(D -1)(D -2) \cdots (D -
2n)}{(2n -1)}\right] \left(\frac{\dot{a}}{a}\right)^{2n} + \mbox{Total
time derivative}
\end{equation}

For a given $D$, the reduced Lagrangian is obtained as,
\begin{eqnarray}
L^D_{\mathrm{reduced}} & := & \sum_{n = 1}^{[D/2]} \alpha_n \lambda^{2n
} \int_{\mathrm{cell}}\sqrt{|g|}L_n^D \nonumber\\
& = & \sum_{n = 1}^{[D/2]} \xi^{1 - D} \alpha_n \lambda^{2n} a^{D -1}
\left[-\frac{(D -1)(D -2) \cdots (D - 2n)}{(2n -1)}\right]
\left(\frac{\dot{a}}{a}\right)^{2n} \nonumber \\
& = & - \left[\xi^{1 - D} a^{D -1} \alpha_1 (D -1)(D -2) \left(\lambda
\frac{\dot{a}}{a}\right)^2\right] \times \nonumber \\ & & \hspace{1.0cm}
\left[ 1 + \sum_{n = 2}^{[D/2]} \frac{\alpha_n}{\alpha_1} \left[\frac{(D
-3)(D -4) \cdots (D - 2n)}{(2n -1)}\right] \left(\lambda
\frac{\dot{a}}{a}\right)^{2n -2} \right] \nonumber \\
& = & - \left[\xi^{1 - D} a^{D -1} \alpha_1 (D -1)(D -2) \left(\lambda
\frac{\dot{a}}{a}\right)^2\right] \times \nonumber \\ & & \hspace{1.0cm}
\left[ 1 + \sum_{n = 1}^{[D/2] - 1} \frac{\alpha_{n + 1}}{\alpha_1}
\left[\frac{(D -3)(D -4) \cdots (D - 2n -2)}{(2n + 1)}\right]
\left(\lambda \frac{\dot{a}}{a}\right)^{2n} \right]
\label{LovelockSeries}
\end{eqnarray}

In the above $\lambda$ is a constant with dimensions of length so that
each of the term in the sum has the same dimension. The
$\alpha_n/\alpha_1$ are arbitrary dimensionless constants and we take
the cell to have the comoving volume given by $\xi^{1 - D}$ where
$\xi^{-1}$ is another length scale. Note that $\alpha_1$ could be a
dimensionful parameter.  

Comparison of the second square brackets in (\ref{EffectiveLagrangian})
and (\ref{LovelockSeries}) suggests the choices for $\lambda$ and
$\alpha_n/\alpha_1$, $n = 1, 2, \cdots , [\Case{D}{2}] - 1$ :
\begin{eqnarray}
x & \leftrightarrow & \lambda\frac{\dot{a}}{a}  ~ \Rightarrow ~ \lambda
:= 2 \Delta \gamma^{3/2}\lP ~,~ \nonumber \\
\frac{\alpha_{n + 1}}{\alpha_1} & \leftrightarrow & \frac{2n + 1}{(D -
3)(D - 4)\cdots (D - 2n -2)} \cdot \frac{(2 n - 1)!!}{(n + 1)! 2^n (2n +
1)} \ .
\end{eqnarray}
Note that $\lambda$ being constant, selects the $\bar{\mu}$-scheme
($\alpha = -1/2)$. 

Matching the powers of $a$ in the first square brackets however requires
$D = 4$ and determines $\alpha_1 = (2\kappa \lambda^2)^{-1}$. But then
the sum over $n$ drops out. There does not seem to be a way to define
any $D \to \infty$ limit such that (a) the finite sum can be extended to
an infinite power series {\em and} (b) the first factors match. 

Thus, although it is possible to get a reduced Lagrangian which depends
only on $a, \dot{a}$, from a Lagrangian with higher powers of
curvatures, this requires Lovelock Lagrangian in arbitrarily high
space-time dimensions to generate the infinite power series in
$\dot{a}/a$. In addition, the first factors do not match. This route for
seeking an interpretation of the quantum corrections summarised in LQC
effective action is not viable.

There is another way to obtain second order equation for the FRW sector
from a general effective action which we discuss in the next section.

\section{General Effective Action} 
In quantum field theory one constructs an effective action, formally,
from the Feynman path integral \cite{BenLee}. In various approximations,
one attempts to compute it.  Weinberg \cite{Weinberg} has given a
general characterization of an effective action ({\em not to be
requantized}) which is supposed to incorporate at least a class of
quantum corrections. One chooses a set of fields, assumes certain
invariances and also {\em locality} in the sense that the action is to
be made up of terms each of which is an integral over positive integer
powers of fields and their derivatives. For a quantized gravity
\footnote{The idea that general relativity can be interpreted as an
effective field theory by introducing higher order curvature invariants
in the original action is discussed elaborately in
ref.\cite{Donoghue}.}, the field would be the metric tensor, its
derivatives would be expressed in terms of the Riemann and Ricci
curvatures and the invariance demanded is the general covariance
\footnote{We could also include covariant derivatives of the Riemann and
Ricci tensors, but for our purposes, these will not be needed.}. Thus
the general form is expected to be a power series in scalars constructed
from the Riemann tensor, the Ricci tensor and the Ricci scalar. One can
also put in the cosmological constant term.  The general action will be
an infinite series in these scalars whose {\em coefficients would depend
on specific underlying quantum theories.} Within this class of effective
actions, different proposals for a fundamental quantum field theory are
distinguished only by these coefficients.

These coefficients are in general dimensionful. Since the Riemann tensor
has length dimension of -2, different powers will have different
dimensions while the action must be dimensionless. This fixes the
dimensions of the coefficients. Observe that quantum gravity provides a
natural length scale, namely the Planck length $\lP$. So one can always
use $\lP$ to convert the coefficients to dimensionless numbers which
encode the specifics of the underlying quantum gravity theory.

Now the observation is that these coefficients are {\em independent of
the field configurations} and one can hope to compare different theories
by specializing to various physical contexts, such as the FRW metric,
the metrics of diagonalised homogeneous models, spherically symmetric
metric etc. In effect, one is carrying out (say) a symmetry reduction
after quantization albeit only incorporating those features which are
captured by the form taken for the general effective action. In
contrast, the LQC effective actions derived above, incorporate a subset
of corrections (the holonomy corrections in the gravitational sector) in
a quantization of {\em a classically reduced theory.} 

The effective action framework thus affords on the one hand a comparison
of different fundamental theories and on the other hand a comparison of
reduction after quantization and quantization after reduction
approaches. These comparisons are of course limited due to inclusion of
only a subset of corrections.  

With these general remarks, let us consider the context of flat FRW
models. 

The locality assumption mentioned above is valid in a perturbative
analysis (splitting the metric in a background and a fluctuation). In
the context of a time dependent scale factor, perturbation would be
appropriate only for a slow variation.  Furthermore, quantum effects
would be expected to be small when the background is `almost classical'
which in the context of flat, isotropic model, corresponds to ``late
time'', $\lP\Case{\dot{a}}{a} \ll 1$.

As can be seen explicitly from the equations (\ref{Riemann},
\ref{RicciTensor}, \ref{RicciScalar}), the scalars constructed from
polynomials in Riemann and Ricci tensors will be polynomials in
$\Case{\ddot{a}}{a}, \Case{\dot{a}}{a}$. If we allowed derivatives of
these tensors, then higher time derivatives would also be present. The
general effective action would then be an infinite series in $H :=
\Case{\dot{a}}{a}$ and its time derivatives (apart from $a^3$ from the
$\sqrt{g}$ factor). Note that the classical action part, modulo a total
time derivatve, has no derivatives of $H$.

At this stage further approximations to the above action are
conceivable. A general higher derivative action will have many more
solutions than those of the leading order (classical) action. The
perturbative nature of the quantum corrections should lead to small
deviations from the classical solutions for self consistency. In
particular, the space of classical solutions should remain the same
(although the individual solutions will of course change). In effect,
this requires the higher derivative terms to be thought of as being
determined by the classical solutions, corrected order-by-order.  The
effective action is then again a polynomial in $H$, although not
manifestly so. This is a correct procedure to interpret the higher
derivative action obtained in a perturbative context \footnote{We thank
an anonymous referee for drawing our attention to this point.}.  Note
that the same arguments also apply for obtaining corrections to the
classical solutions from the LQC effective action. However, the LQC
effective action being a function of only $H$, one can keep this
procedure implicitly understood.

Recall that the LQC Lagrangian of equation (\ref{EffectiveLagrangian}),
is only the leading order of the WKB method and hence has only the
classical degrees of freedom visible. The higher order corrections from
the WKB, will introduce higher time derivatives as well which however
are not yet available. In effect, one has implicitly {\em dropped the
higher derivative corrections coming from LQC}. Had these terms been
available in LQC effective action, one would apply the same
considerations as given above.

At the present state of availability of quantum corrections from LQC, it
seems more `fair' to approximate the full theory effective action by
explicitly dropping all terms containing derivatives of $H$, as has been
implicitly done for the LQC effective action. Now both actions have same
form and comparison of coefficients is possible.

To summarise, there are two ways to compare the two effective actions
depending upon the form in which they are available. If both effective
actions have higher derivatives, then in each one, the higher
derivatives can be treated as being determined by solutions of the lower
order equation of motion. In effect, one can compare the solutions
connected to the {\em same} classical solution. Alternatively, if one
action has no higher derivative terms, then the second one can be
brought to the same form by dropping the higher derivative terms. Now
the actions themselves can be compared directly.

Either of these is a possible method by which one can compare the LQC
effective Lagrangian with an effective Lagrangian constructed for the
full theory in any particular version of quantum gravity. It is also
possible to carry out a similar comparison when effective Lagrangians
become available for anisotropic LQC, spherically symmetric models etc.
The full theory effective action will have the {\em same} coefficients
in all these cases.

Here we note another point relevant for comparisons. Suppose an
effective action is given as a series in curvature scalars with certain
specific coefficients, $L \sim c_1R + c_{2,0}R^2 + c_{2,1}R_{ab}R^{ab} +
c_{2,2} R_{abcd}R^{abcd} + \cdots $ which is a function of $\dot{H}$ and
$H$. In either of the methods described above, each of the curvature
scalars will effectively be a monomial in $H$ with the power determined
by the dimensional consideration and the coefficient determined by
actual computation. These are fixed coefficients independent of the
quantum theory. For example, we could get $R^2 \approx k_{2,0} H^4,
R_{ab}R^{ab} \approx k_{2,1} H^4, R_{abcd}R^{abcd} \approx k_{2,2} H^4$
and so on. The net result will be a power series in $H$ whose
coefficients will be combinations of $c_{m,n}$ and
$k_{m,n}$\footnote{The combinations of these coefficients will in
general be different in the two methods.}. The $H^4$ coefficient for
instance would be $(c_{2,0}k_{2,0} + c_{2,1}k_{2,1} + c_{2,2}k_{2,2})$.
If we were to attempt inferring the theory dependent coefficients
$c_{m,n}$ by a comparison, then the FRW sector can at best yield some
constraints on the combinations and other (less symmetric) sectors will
be needed. 

\section{summary and conclusions}

In this work, we first obtained the effective Lagrangian(s) for the
gravitational sector of isotropic LQC. The domain of validity of the
effective Hamiltonian (and hence the Lagrangian) is the regime
$\lP|\dot{a}/a| \ll 1$. This is given as an infinite power series. There
are two ways to obtain such series from an effective action for the full
theory: (a) using actions whose degrees of freedom exactly match with
the classical ones (same order of equations of motion), eg the Lovelock
actions and (b) invoking a suitable approximation to restrict to the
classical degrees of freedom as the dominant/relevant ones (directly by
dropping higher derivative terms or indirectly by treating higher
derivatives as being determined by lower order solutions). The former
however requires considering arbitrarily high space-time dimensions and
does not yield a form consistent with the LQC effective action. The
latter, though it involves an approximation, is more general and
consistent with the domain of validity of the LQC effective action as
well as with the nature of quantum corrections implicit in the higher
curvature action.  This naturally restricts the LQC effective action to
the $\bar{\mu}$-scheme. We would like to note that a comparison with a
greater precision (i.e. without dropping higher derivative terms) will
be possible if the effective action for LQC could be computed including
higher time derivatives of the scale factor.  However, in order to
compare different underlying quantum theories, the homogeneous and
isotropic sector alone cannot be sufficient since only certain
combinations of the $c_{m,n}$'s can get constrained. 

The proposed approach of comparing different quantum theories at the
level of effective actions (really at the level of equations of motion
since we do not worry about total derivative terms) is a preliminary one
and can be quantitatively useful only when the effective actions at both
the full and the reduced level are independently and reliably
computable. In the absence of availability of such actions, one can at
best proceed with a qualitative comparison.  

If effective actions are available for different classical sectors, then
the demand that these be obtained from corresponding reductions from a
{\em common} full theory effective action, would be restrictive. This
provides a motivation for obtaining effective actions for several
different classically reduced models \footnote{While this work was being
written up, the eprint of \cite{Thomas} appeared which has some overlap
with our results. Our perspective and approach is however different. We
focus on the effective action for the gravitational sector, discuss the
Lovelock actions as a candidate and point out the appropriateness of
$\bar{\mu}$-scheme.  We do not discuss the matter sector which is
addressed in the cited work.}.

\acknowledgments
Discussions with Kinjal Banerjee, Martin Bojowald, Romesh Kaul and Alok
Laddha are gratefully acknowledged. 
%

%%%%%%%%%%%%%%%%%%%%%%%%%%%%

\end{document}